\documentclass[aps,prd,showkeys,showpacs,nofootinbib]{revtex4}
\global\parskip 6pt

\normalsize

\def\br{\begin{thebibliography}{abc}}
\def\er{\end{thebibliography}}

\begin{document}

\newcommand{\be}{\begin{equation}} \newcommand{\ee}{\end{equation}}
\newcommand{\bea}{\begin{eqnarray}}\newcommand{\eea}{\end{eqnarray}}

\begin{flushright}
\hfill{SINP-TNP/09-26}\\
\end{flushright}
\vspace*{1cm}
\thispagestyle{empty}
\centerline{\large\bf Constraints on the quantum gravity scale from
$\kappa$ -Minkowski spacetime }
\bigskip
\author{A. Borowiec \footnote {borow@ift.uni.wroc.pl}}
\affiliation{
Institute for Theoretical Physics, University of Wroclaw,
pl. Maxa Borna 9, 50-204 Wroclaw, Poland}

\author{Kumar S. Gupta \footnote {kumars.gupta@saha.ac.in}}

\affiliation{Theory Division, Saha Institute of Nuclear Physics, 1/AF
Bidhannagar, Calcutta 700064, India}
\author{S. Meljanac \footnote {meljanac@irb.hr}}

\affiliation{Rudjer Bo\v{s}kovi\'c Institute, Bijeni\v cka  c.54, HR-10002
Zagreb, Croatia}

\author{A. Pacho{\l} \footnote {anna.pachol@ift.uni.wroc.pl}}
\affiliation{
Institute for Theoretical Physics, University of Wroclaw,
pl. Maxa Borna 9, 50-204 Wroclaw, Poland}

\vskip.5cm

\begin{abstract}

We compare two versions of deformed dispersion relations (energy vs momenta and momenta vs energy) and the corresponding time delay up to the second order accuracy in the quantum gravity scale (deformation parameter). A general framework describing modified dispersion relations and time delay with respect to different noncommutative $\kappa$ -Minkowski spacetime realizations is firstly proposed here and it covers all the cases introduced in the literature.
It is shown that some of the realizations provide certain bounds on quadratic corrections, i.e. on quantum gravity scale, but it is not excluded in our framework that quantum gravity scale is the Planck scale. We also show how the coefficients in the dispersion relations can be obtained through a multiparameter fit of the gamma ray burst (GRB) data.

\end{abstract}



\maketitle


\section{Introduction}
Understanding the properties of matter and spacetime at the Planck scale
remains a major challenge in theoretical physics. In spite of several
theoretical candidates, the corresponding experimental data are very limited.
Among the data that is presently available, the difference in arrival time
of photons with different energies from GRB's, as observed by the Fermi
gamma-ray telescope \cite{Fermi,ref3,recent}, may contain information about the
structure of spacetime at the Planck scale \cite{ellis1,ellis2}, see \cite%
{smolin1} for a recent review. There have been attempts to analyze this data
from the GRB's based on the framework of doubly special relativity (DSR)
\cite{dsr1,dsr2,smolin1,smolin2,smolin3}. It was argued \cite{majid1} that the dispersion relation following
from DSR is consistent with the difference in arrival time of photons with
different energies. Similar dispersion relations have also been proposed in
other scenarios, see, e.g., \cite{smolin2, smolin3,mattingly,girelli}, most of which lead to Lorentz symmetry
violation.

In a related development it was found that the dispersion relations
following from the $\kappa$-Minkowski spacetime \cite{k1}-\cite{MR} can be used to analyze the astrophysical data from the GRB \cite{majid1}. In addition, one possible description of the symmetry structure of the DSR is through the Hopf-algebra of the $\kappa$-Minkowski spacetime
\cite{dsr1,glik1,glik2,glik3,boro1,lee1,lee2,rim,ref1}, although alternative descriptions are also possible (see e.g. \cite{ref1,ref2}).
The noncommutative geometry defined by the $\kappa$-Minkowski spacetime can also
be obtained from the combined analysis of special relativity and the quantum
uncertainty principle \cite{sergio1,sergio2}. These observations indicate
that the $\kappa$-Minkowski spacetime could be a possible candidate to
describe certain aspects of physics at the Planck scale.

Recently it has been emphasized that the dispersion relation compatible with
the GRB data is likely to arise from a DSR model where the transformation
laws are changed but the Lorentz symmetry is kept undeformed \cite{smolin1}. Assuming that the DSR symmetry can be described by the $\kappa$-Minkowski Hopf algebra, this implies that the corresponding $\kappa$-Minkowski spacetime should have an undeformed Lorentz algebra but may have a deformed co-algebra. Therefore we identify deformation parameter $\kappa$ with the quantum gravity scale since it appears  as an invariant scale in DSR theory interpretation, dispersion relations become deformed and quantum gravity corrections appear.
It should be mentioned that nonlocality of DSR formalism was recently under debate \cite{DSRdebate}, however recent research points out that there is
no disagreement between the principles of DSR and the observation \cite{DSRdebate2}.
The characterization of nonlocality has been shown to be inapplicable to DSR frameworks based on $\kappa$-Poincare \cite{DSRdebate2}.

In this Letter we shall give a general description of the dispersion
relations arising from $\kappa$-Minkowski spacetime within a class of
realizations where the Lorentz algebra is kept undeformed but the
corresponding co-algebra is deformed \cite{s1,s2,s3,s4,jord,s5}.
We consider two equivalent types of dispersion relations which are characterized by a set of parameters which
can be determined by the choice of the realization. The general framework, firstly proposed here, includes most of
the cases introduced in the literature so far.
 Although the parameters in the dispersion
relations can be calculated for any given choice of the realization, here we
take the point of view that they should be determined using the empirical
data from the astrophysical sources. Identifying deformation parameter with quantum gravity scale, experimental data 
\cite{Fermi,ref3,recent} gives the relation $\frac{M_Q}{M_{Pl}}>1.2$, which indicates the possibility of truly Planckian effects. In our framework, however it is possible to even get
$M_Q=M_{Pl}$, due to proper choice of proportionality coefficient. Besides leading term - proportional to
$\frac{1}{M_{Pl}}$, we consider also quadratic corrections, which might provide a better fit to astronomical data, and have a higher accuracy on bounds of the quantum gravity scale.

In the second section, we discuss the generalized dispersion relations that follow
from our analysis of the $\kappa$-Minkowski Hopf algebra. In the third section we
introduce all the realizations for $\kappa$-Minkowski spacetime coordinates and provide a comparison with recent experimental data which allows us to obtain bounds on quantum gravity scale or model parameters. The fourth
section concludes the paper with some discussions.

\section{Dispersion relations}

Consider the $\kappa$-deformed Minkowski spacetime provided with noncommutative
coordinates $\hat{x}_{\mu}$, where\\ $\mu = 0,1,..,n-1 $
\be
[\hat{x}_{\mu},\hat{x}_{\nu}]  = i(a_\mu \hat{x}_{\nu} - a_\nu \hat{x}%
_{\mu})\ee
and supplemented by the undeformed Lorentz sector:
\be
\left [ M_{\mu \nu}, \hat{x}_{\lambda} \right ] = \hat{x}_{\mu} \eta_{\nu
\lambda} - \hat{x}_{\nu} \eta_{\mu \lambda} - i(a_\mu M_{\nu \lambda} -
a_\nu M_{\mu \lambda})\ee\be
\left [ M_{\mu \nu},M_{\lambda \rho} \right ] = M_{\mu \rho} \eta_{\nu
\lambda} - M_{\nu \rho} \eta_{\mu \lambda} + M_{\nu \lambda} \eta_{\mu \rho}
- M_{\nu \lambda} \eta_{\nu \rho}\ee
where $a_\mu$ is a Lorentz vector and $\eta_{\mu \nu} = \mathrm{diag}%
(-1,1,...,1)$. The quantity $a^2 = a_\alpha a^\alpha$ is Lorentz invariant
and $\kappa^2 \equiv \frac{1}{a^2}$. The above Lie algebra satisfies the
Jacobi relations and in the limit $a_\mu \rightarrow 0$ the commutative
space with the usual action of the Poincare algebra is recovered.
We introduce momenta $P_{\mu}$ transforming vector-like under the Lorentz
algebra \be\left [ P_{\mu},P_{\nu} \right ]=0, \ee\be
\left [ M_{\mu\nu},P_{\lambda} \right ]= P_{\mu} \eta_{\nu\lambda}-
P_{\nu} \eta_{\mu\lambda}\ee
in order to form undeformed Poincar\'{e} algebra.

It is worth noticing that, due to Jacobi identity, canonical commutation relations: $
[P_{\mu },\hat{x}%
_{\nu }]=\eta _{\mu \nu }$
 are not satisfied. In fact, one has, instead, the deformed Weyl algebra as follows:
\begin{equation}
\lbrack P_{k},\hat{x}_{0}]=0,\qquad \lbrack P_{k},\hat{x}_{j}]= \imath \delta
_{jk}\left(a_0 P_{0}- \sqrt{1-a^{2}P_{\mu }P^{\mu }}\right)
\label{L7}
\end{equation}%
\begin{equation}
\lbrack P_{0},\hat{x}_{j}]=a_0P_{j},\qquad \lbrack P_{0},\hat{x}_{0}]=\imath
\sqrt{1-a^{2}P_{\mu }P^{\mu }}  \label{L8}
\end{equation}%
(see, e.g., \cite{s1, jord}). These relations close the algebra (1)-(5) and modify  Heisenberg uncertainty  relations (see, e.g., \cite{Liberati} and references therein). It appears that interesting class of representations of the algebra
(1)-(7) can be induced from representations of the deformed spacetime algebra itself (1) (see \cite{s1} - \cite{s5} for details).
We shall call the algebra (1)-(7) a DSR algebra, as proposed in \cite{JKG}, since its different
realizations lead to different doubly (or deformed) special relativity models with
different physics encoded in deformed dispersion relations. Let us clarify
this point in more detail. In terms of momenta $P_{\mu }$, we have
undeformed dispersion relations given by the standard Poincar\'{e} Casimir
operator\begin{equation}
P^{2}+m_{ph}^{2}=0  \label{nondef}
\end{equation}%
where $m_{ph}$ is the physical mass, which comes from the representation of the
Poincar\'{e} algebra. However, the standard Casimir operator $P^2$ does not satisfy:
\be [P^{2},\hat{x}_{\mu }]=2P_{\mu }\ee
This motivates us to look for another invariant
operator - the deformed Casimir operator $\mathcal{C_{\kappa }}^{2}$ for which~:
\be
[M_{\mu \nu },\mathcal{C_{\kappa }}^{2}]=\lbrack \mathcal{C_{\kappa }}%
^{2},P_{\mu }]=0;\quad \lbrack \mathcal{C_{\kappa }}^{2},\hat{x}_{\mu
}]=2P_{\mu }\ee
It leads to the deformed dispersion relation
\begin{equation}
\mathcal{C_{\kappa }}^{2}+m_{\kappa }^{2}=0  \label{def}
\end{equation}
with the deformed mass parameter $m_{\kappa }$. It turns out that in the most
concrete realizations the interrelation between this two invariants has the
following form \cite{s1, jord}
:\be P^{2}=\mathcal{C_{\kappa }}^{2}(1+\frac{a^{2}}{4}\mathcal{C_{\kappa }}^{2})\ee
Particularly \be m_{ph}^{2}=m_{\kappa }^{2}(1-\frac{a^{2}}{4}m_{\kappa }^{2})\ee
Therefore for photons: $m_{ph}=m_{\kappa }=0$ and, as a consequence, dispersion relations obtained from (\ref{nondef}) and (\ref{def}) are identical.
One can, however, see that in general both expressions  have
the same classical limit $a_\mu \rightarrow 0$ but differ by order as
polynomials in $a$. It happens that both dispersion relations (\ref{nondef},\ref{def}) can
be rearranged into the convenient form:
\begin{equation}  \label{disp1}
E \simeq |{\vec{p}}|\frac{G_{1}}{G_{2}}+\frac{m^{2}}{2|{\vec{p}}|}
\end{equation}
\begin{equation}  \label{disp2}
\frac{G_{1}}{G_{2}} \simeq 1+c_{1}\frac{|{\vec{p}}|}{M_{Q}}+c_{2}\frac{|{%
\vec{p}}|^{2}}{M_{Q}^{2}}
\end{equation}
if one restricts oneself only to the second-order accuracy and introduces locally measurable momenta $(E,\vec{p})$.
Here $G_{1}$ and $G_{2}$ are model-dependent functions of $\frac{|\vec{p}|}{M_{Q}%
}$ and $\frac{m^{2}}{M_{Q}^{2}}$ satisfying the conditions $%
G_{1}(0)=G_{2}(0)=1$, and $c_{1}$ and $c_{2}$ are model-dependent constants.
Taking $p_{0}=E$ and $a^{\mu}=(M_{Q}^{-1},0,0,0)$ as a time-like vector so that $a^{2}=-M_{Q}^{-2}<0$ where $M_{Q}$ is the quantum gravity scale and assuming that $m<<E<<M_{Q}$, we have energy dependence up to second order in $\frac{1}{M_{Q}^{2}}$.
Eqs. (\ref{disp1}) and (\ref{disp2}) provide the general dispersion
relations within the $\kappa$-Minkowski Hopf algebra framework and up to the
second order in $\frac{1}{M_Q^2}$. These constants, $c_1$ and $c_2$, are
additional parameters appearing in the dispersion relation and time delay
formula. It has been observed in \cite{smolin1} that a proper analysis of
the GRB data using dispersion relations may require more than just the
parameter given by the quantum gravity scale $M_Q$. In fact, in order to diversify between different DSR models (see below) one has to assume some numerical value for the parameter $M_Q$ (e.g. $M_Q=M_{Pl}=1,2\times 10^{19} GeV$) and fit the values for linear $c_1$ as well as quadratic $c_2$ corrections \cite{Albert,Aharonian}. Otherwise one can choose parameters $c_1$ and $c_2$ from the quantum gravity model and experimentally fit the quantum gravity scale $M_Q$.

 In our formalism, the
parameters $c_1$ and $c_2$ arise naturally in the dispersion relations,
which can be obtained by fitting with the empirical data. They can also be
calculated theoretically once the choice of the realization of the $\kappa$%
-Minkowski Hopf algebra is fixed. In this sense, the parameters $c_1$ and $%
c_2$ incorporate the information of the quantum gravity vacuum, analogous to
certain fuzzy descriptions of quantum gravity \cite{f1,f2}.\newline
Alternatively one can consider dispersion relation in the form:
\begin{equation}  \label{disp3}
|\vec{p}|\simeq E \left(1-b_1\frac{E}{M_Q}+b_2\frac{E^2}{M^2_Q}\right)
\end{equation}
which coincides with (\ref{disp1}) and (\ref{disp2}) up to the second order accuracy
provided $c_1=b_1$ and $c_2=2b_1^2-b_2$, i.e.\be
E\simeq |\vec{p}| \left(1+b_1\frac{|{\vec{p}}|}{M_Q}+(2b_1^2-b_2)\frac{|{%
\vec{p}}|^2}{M^2_Q}\right)\ee
This gives time delay formula as: 
\be  \label{td}
\Delta t\simeq -\frac{l}{c}\frac{|\vec{p}|}{M_Q}\left(B_1+\frac{|\vec{p}|}{M_Q%
}B_2\right) =\nonumber\ee
\be
 -\frac{l}{c}\frac{E}{M_Q}\left(2b_1-3b_2\frac{E}{M_Q}\right)
\ee
where $B_1=2c_1=2b_1,\quad B_2=c_2-4c_1^2=2b_1^2-3b_2$ and $l$ is a distance
from the source of high-energy photons. This last equations might  be
more suitable for our purposes since photon energy is a well measurable
physical quantity. The case $M_Q=M_{Pl}$ is not excluded, thus natural relation between the quantum gravity scale and the Planck scale arises here (see below for details).
 In general, due to Lorenz Invariance Violation (LIV): $|\vec{p}|\neq E$, the second-order contributions to "momentum" and "energy" versions of the time delay formula (\ref{td}) are different. However, for $b_1=0$, one has $B_2=-3b_2$ and both contributions are the same.
The formula (\ref{td}) describes absolute time delay between
deformed and undeformed cases. What one really needs, in order to compare against experimental data, is the relative time delay i.e. time lag between two photons with different energies $E_{l}<E_{h}; \delta E=E_{l}-E_{h}; \delta E^2=E^2_{l}-E^2_{h}$ (with lower-energy photons $E_l$ arriving
earlier).
\begin{equation}
\Delta \delta t\simeq - \frac{l}{c}\left( 2b_{1}\frac{\delta E}{M_{Q}}-3b_{2}%
\frac{\delta E^{2}}{M_{Q}^{2}}\right)
\end{equation}%
Following the lines of thoughts presented in \cite{jacob,xiao} one can also take into account
the universe cosmological expansion: for the photons coming from a redshift $z$ one gets
\begin{equation}  \label{cosm}
\Delta \delta t\simeq -2b_1\frac{\delta E_e}{M_{Q}}%
\int_{0}^{z}\frac{1+z^{\prime }}{h(z^{\prime })}dz^{\prime }+3b_2\frac{%
\delta E_e^{2}}{M_{Q}^{2}}\int_{0}^{z}\frac{(1+z^{\prime })^{2}}{%
h(z^{\prime })}dz^{\prime }
\end{equation}%
where: $h(z^{\prime })=H_{0}\sqrt{\Omega _{\Lambda }+\Omega _{M}(1+z)^{3}}$
and $H_{0}=71/s/Mpc$ is the Hubble parameter, $\Omega _{M}=0.27$ - the matter density and $%
\Omega _{\Lambda }=0.73$ - the vacuum energy density are cosmological parameters represented by their present-day values:
$E_e$ denotes the redshifted photon's energy as measured on Earth.

\section{Time delay formulae for different realizations of DSR algebra}

The DSR algebra (1)-(7) admits a wide range of realizations:\be
\hat{x}_{\mu } =x^{\alpha }\phi _{\alpha \mu }(p; M_Q ), \quad
M_{\mu \nu } =x_{\alpha }\Gamma _{\mu \nu }^{\alpha }(p; M_Q ),\nonumber\ee
\be P_{\mu } =\Lambda _{\mu }(p; M_Q )\ee
in terms of (undeformed) Weyl algebra \footnote{For details on relation between deformed and undeformed Weyl algebra see \cite{ABAP4}, particularly Proposition 3.1 and 3.3.} satisfying the canonical commutation relations:\be
\left[ p_{\mu },x_{\nu }\right]=-i\eta _{\mu \nu }, \quad
\left[ x_{\mu },x_{\nu }\right]=0, \quad
\left[ p_{\mu },p_{\nu }\right]=0.
\ee
with proper classical limit: $\hat{x}_{\mu } =x_\mu,\quad M_{\mu \nu }=x_\mu p_\nu-x_\nu p_\mu,\quad P_\mu=p_\mu$ as $M_Q\rightarrow\infty$.
It may be noted that there are infinitely many choices for $P_{\mu }$
compatible with Lie algebra, eqs. (1)-(3). In these realizations, $x_{\mu }$
and $p_{\nu }$ do not transform as vectors under the action of $%
M_{\mu \nu }$. They provide (commuting) position $x^\mu$ operators and local, physically measurable, momentum $p_{\mu }=-i\partial_{\mu }$ which define measurable frame  for DSR theories, for discussion of DSR phenomenology, see \cite{Liberati} and references therein.\newline

\textbf{Noncovariant realizations} cover a huge family of DSR-type  models generated by two arbitrary analytic
functions $\psi ,\gamma $:
\be \hat{x}^{i}=x^{i}\phi (A),\quad \hat{x}^{0}=x^{0}\psi (A)+\imath ax^{k}\partial _{k}\gamma (A)\ee
where $A=-ap=-\frac{E}{M_{Q}}$ and $\phi (A)=\exp \left( \int_{0}^{A}\frac{%
(\gamma (A^{\prime })-1)dA^{\prime }}{\psi (A^{\prime })}\right)$; $\psi (0)=1$ \cite{s1}, \cite{jord}.
Photon's dispersion relations with respect to both undeformed and deformed Casimir operators (\ref{nondef}, \ref{def})
can be recast as follows:
\begin{equation}\label{a2}
|\vec{p}|=E\frac{1-\exp \left( -\int_{0}^{A}\frac{%
dA^{\prime }}{\psi (A^{\prime })}\right)}{A}
\exp \left( \int_{0}^{A}\frac{
\gamma (A^{\prime })dA^{\prime }}{\psi (A^{\prime })}\right)
\end{equation}
Recent experimental data disfavour models with only linear accuracy.
In order to calculate second-order contribution
stemming from (\ref{a2}) one needs the following expansion:
\begin{equation}
\psi =1+\psi _{1}A+\psi _{2}A^{2}+o(A^{3});\quad \gamma =\gamma _{0}+\gamma
_{1}A+o(A^{2})
\end{equation}%
This provides general formulae for the coefficients $b_{1},b_{2}$ in (\ref%
{disp3}) and $B_{1},B_{2}$ in (\ref{td},\ref{cosm}):
\be b_{1}=\frac{1}{2}(2\gamma _{0}-1-\psi _{1})\ee
\be b_{2} =\frac{1}{6}(1+3\psi _{1}+2\psi _{1}^{2}-\psi _{2}+3\gamma
_{0}^{2}-3\gamma _{0}+3\gamma _{1}-6\gamma _{0}\psi _{1})\ee
\be B_{1} =2\gamma _{0}-1-\psi _{1};\ee
\be B_{2} =\frac{1}{2}(\gamma
_{0}^{2}-\psi _{1}^{2}-\gamma _{0}+2\psi _{1}\gamma _{0}-\psi _{1}+\psi
_{2}-3\gamma _{1})\ee
The general case of Hermitian (Hilbert space) realization \cite{jord} requires (in physical dimension four): $\psi ^{\prime
}+3\gamma=0$ which corresponds to: $\psi _{1}=-3\gamma _{0};\psi _{2}=-\frac{3%
}{2}\gamma _{1}$ and gives rise to formulas:
\be
b_{1}=\frac{1}{2}(5\gamma _{0}-1)\ee
\be b_{2}=\frac{1}{6}(1+39\gamma _{0}^{2}-12\gamma _{0}+\frac{9}{2}\gamma _{1})\ee
\be B_{1}=5\gamma _{0}-1;\quad B_{2}=\frac{1}{2}(4\gamma _{0}^{2}+2\gamma _{0}-%
\frac{9}{2}\gamma _{1})\ee
Particularly, for $\gamma _{0}=\frac{1}{5}$, one can reach $b_{1}=0$ and
\be b_{2}=\frac{3}{4}\gamma _{1}+\frac{2}{75}\ee
a) Non-covariant realizations contain representations generated by the
\textbf{Jordanian one-parameter family of Drinfeld twists} (for details see
\cite{jord}): $\psi =1+rA, \gamma=0$, hence $\psi _{1}=r,\psi_2 =\gamma
_{0}=\gamma _{1}=0$.\newline
The corresponding time delay coefficients for photons are:\be
b_{1}=-\frac{1}{2}(1+r);\quad b_{2}=\frac{1}{6}(1+3r+2r^{2});\quad
B_{2}=-\frac{1}{2}r(r+1)\ee
so one gets an upper-bound $B_2\leq -\frac{3}{8} $. However, in (\ref{disp3}) one gets $b_2 \geq -\frac{1}{4}$ which provides a lower-bound for $b_2$.
Particularly, for $r=-1$ one gets $b_{1}=0=b_2$. This
corresponds to Poincar\'{e}-Weyl algebra \cite{jord} and provides no time delay for photons.
A particular Jordanian Hermitian  case in $n=4$ dimensions requires $r=3$ and gives \be
\Delta t=\frac{l}{c}\frac{|\vec{p}|}{M_{Q}}\left( 4+3\frac{|\vec{p}|}{M_{Q}}%
\right)\ee  and leads to "time advance".
Another case of dispersion relation worth to consider:
\be |\vec{p}|=\frac{E}{1-\frac{E}{M_Q}} \ee
is recovered here for parameter $r=1$, and one gets $b_1=-1; b_2=1$ in time delay formulas. This case differs by sign from the dispersion relation considered in \cite{smolin3}.
Nevertheless, the original formulae proposed there:
\be \frac{E^2}{(1+\frac{E}{M_Q})^2}-|\vec{p}|^2=m^2\ee
 can be also reconstruct within this formalism for the following choice: $\psi=1-3A+2A^2;\gamma=0$ and one gets $b_1=b_2=1$ in time delay for photons. However, we still do not know if this realization is possible to obtain by Drinfeld twist.\\
b) The $\kappa -$Minkowski spacetime can be also implemented by
one-parameter family of \textbf{Abelian twists} \cite{jord,lee1,lee2}.
Abelian twists give rise to : $\psi =1, \gamma =s= \gamma _{0}, \gamma _{1}=\psi_1=\psi_2=0$ and
\be b_{1}=\frac{1}{2}(2s-1);\quad b_{2}=\frac{1}{6}(3s^{2}-3s+1)\ee
\be \Delta t=-\frac{l}{c}\frac{|\vec{p}|}{M_{Q}}\left( 2s-1+\frac{|\vec{p}|}{%
2M_{Q}}s(s-1)\right)\ee
so $B_2\geq \frac{1}{8}$ provides bounds from below ($s=\frac{1}{2}$). And analogously one obtains an upper-bound for $b_2\leq \frac{1}{6}$ in (\ref{disp3}).
The case $b_{1}=0$ gives  $\Delta t\simeq -\frac{l}{c}\frac{E^{2}}{8M_{Q}^{2}}$. Moreover, the Hermitian realization requires $s=0$ and provides $\Delta t=+\frac{l}{c}\frac{|\vec{p}|}{M_Q}$, "time advance" instead. (The corresponding dispersion relation has been also found in \cite{LZ}.)
As has been shown in \cite{boro1,jord} the case $s=1$
reproduces the standard DSR theory \cite{smolin1}-\cite{dsr2}, \cite{glik1}-\cite{glik3} which is related with the so-called bicrossproduct basis
\cite{MR}. The time delay formula taking into account the second-order contribution reads now as
\be\label{dsr}
\Delta t=-\frac{l}{c}\frac{|\vec{p}|}{M_{Q}}=-\frac{l}{c}\frac{E}{M_{Q}}\left(1-\frac{E}{2M_{Q}}\right)\ee
Assuming $M_Q=M_{Pl}$, the leading term of (\ref{td}) is $\frac{2b_1}{M_{Pl}}$, which compared with recent results, $\frac{M_{Q}}{M_{Pl}}>1.2$, gives $b_1<0,417$. Particularly, for Jordanian realizations one obtains a lower bound for parameter $r>-1.208$, analogously for Abelian realizations $s<0.604$ which provides upper bound.
Finally, it might be of some interest in physics to study models with first-order corrections
vanishing, i.e. $b_{1}=0$. This is due to the fact that it is unlikely to observe LIV at the linear order of $\frac{E}{M_{Q}}$.
This yields $ \psi _{1}=2\gamma _{0}-1$\be
b_{2}=\frac{1}{6}(\gamma _{0} -\psi _{2}+3\gamma _{1}-\gamma _{0}^{2})\ee
\be B_{2}=-3b_{2}=-\frac{1}{2}(\gamma _{0} -\psi _{2}+3\gamma _{1}-\gamma
_{0}^{2})\ee
\textbf{General covariant realization} \cite{s2}.
One can distinguish the interesting case of covariant realizations of
noncommutative coordinates:\newline
\be \hat{x}_{\mu }=x_{\mu }\phi +i(ax)\partial _{\mu }+i(x\partial )(a_{\mu
}\gamma _{1}+ia^{2}\partial _{\mu }\gamma _{2})\ee
with the dispersion relation \be
\ m_{ph}^{2}=\frac{E^{2}-\vec{p}^{2}}{(\phi -\frac{E}{M_Q})^{2}-\frac{%
\vec{p}^{2}}{M^2_Q}}\ee
with respect to (\ref{nondef}), which does not yield time delay for photons.
It is worth noticing, however, that for the special 
choice of $\phi =1$ the last formula recovers Magueijo-Smolin type of covariant dispersion relations (DSR2),
see ref. \cite{smolin3}, formula (3).\newline
One can consider the second Casimir operator in this realization:
$\mathcal{C_{\kappa }}^{2}=\frac{2}{M_Q^{2}}\left( \sqrt{1+M_Q^{2}\frac{p^{2}%
}{\left( \phi +\frac{p_0}{M_Q}\right) ^{2}+\frac{p^{2}}{M^2_Q}}}-1\right)$
but there is no time delay either.\newline

\textbf{Comparison against experimental data}\\
All the above results can be directly translated and compared with recent numerical data obtained by MAGIC, FERMI or other space or ground experiments. In the following we shall compare data given in \cite{Albert,Aharonian} within our framework.
Our formula for time delay (\ref{td}) coincides with the one introduced in \cite{Aharonian} for change of speed of light due to quantum gravity corrections.
\be\label{c}
c^{\prime }=c\left( 1+\xi \frac{E}{M_{Q}}+\zeta \frac{E^{2}}{M_{Q}^{2}}%
\right)=\nonumber\ee
\be
c\left( 1+2b_{1}\frac{E}{M_{Q}}+(4b_{1}^{2}-3b_{2}) \frac{E^{2}}{M_{Q}^{2}}
\right)\ee

First let us consider model-independent limits on the quantum gravity scale based on astrophysical data. Let us assume, as in \cite{Aharonian}, (in general it is not necessary in our models as mentioned in previous section) that quantum gravity scale is Planck mass: $M_{Q}=E_{P}=1.22\times 10^{19}GeV$. For linear corrections in speed of photons (with quadratic corrections vanishing $\zeta=0$, or equivalently $b_2=\frac{4}{3}b_1^2$ in (\ref{c})) one obtains the following bounds for coefficients $b_1,b_2$ in the time delay formula (\ref{td}):\\
- for GRB's $|b_{1}|<35-75$;\\
-for active galaxies (Whipple collaboration during flare at 1996) $|b_{1}|<100$;\\
-MAGIC (2005) $|b_{1}|\approx 15$;
with limits $ |b_{1}|<30$;\\
-and if $c^{\prime }$ is helicity dependent $ |b_{1}|<0.5\times 10^{-7}$.\\
It should be noticed that results simultaneously constraining linear and quadratic corrections are very rough, e.g. $|\xi|<60$ and $|\zeta|<2.2\times 10^{17}$ \cite{Albert,Aharonian} gives rise to $|b_1|<30$ and $1200-0.73\times 10^{17}<|b_2|<1200+0.73\times 10^{17}$.
More experimental data one can find in \cite{Fermi,Albert,Aharonian} and references therein.

Moreover, one can consider model-dependent limits on the quantum gravity scale, since coefficients $b_1,b_2$ are connected with the above introduced, twist realizations of $\kappa$ -Minkowski spacetime. We can choose the type of model (e.g., Hermitian Jordanian, Abelian) and provide the bounds for the quantum gravity scale without $M_Q=M_P$ assumption.\\ Assuming
\textbf{Jordanian} realization in Hermitian case (with $\zeta=0$) the quantum gravity scale bounds are:\\
$M_{Q} >28\times 10^{17} GeV$ (for the limit obtained by MCCF method); \\
$M_{Q}>20,8\times 10^{17}GeV$ (from the wavelet analysis). \\
Taking into account only quadratic corrections in this case we do not obtain any bound for $M_Q$ scale due to $r=-1$ case discussed in the previous section ($\xi=0\Rightarrow b_1=0\Rightarrow b_2=0$).
Analogously, one can do this analysis for
\textbf{Abelian twists}.
For the linear correction in \textbf{Abelian Hermitian} realization we obtain the following bounds on the quantum gravity scale: $M_{Q}>7.2\times 10^{17} GeV $ (the limit obtained by MCCF method). Limit obtained from wavelet analysis: $M_{Q}>5.2\times 10^{17} GeV$.
                                                                                                                                                                                                                                                                                                    \textbf{Abelian DSR} realization provides exactly the same bounds.
However considering only quadratic corrections in Abelian, no longer Hermitian neither DSR case, ($\xi=0\Leftrightarrow s=\frac{1}{2}$) one obtains: $ M_{Q}>1.31\times 10^{9} GeV$ (within the MCCF method).
One can notice that all bounds on the quantum gravity scale are lower bounds.

\section{Discussions}
We have shown that dispersion relations arising from our analysis of the $\kappa$-Minkowski spacetime contain multiple parameters, which depend on the choice of the realization and deformation parameter which is identified with the quantum gravity scale. While the pure linear suppression by the quantum gravity scale is theoretically allowed, its exact form
is not yet established from the recent analysis of the GRB data \cite{Fermi,ref3,recent}. However, our analysis predicts a more general form of the in vacuo dispersion relations (\ref{disp3}), which typically contain terms both linearly and quadratically suppressed by the quantum gravity scale $M_Q$. Moreover, the dispersion relations obtained here contain parameters which depend on the choice of the realizations of the $\kappa$-Minkowski algebra. A priori there is no basis to prefer one realization over another, which should be an empirical issue. It would therefore be best to obtain all the parameters in the dispersion relations from a multiparameter fit of the GRB data. We believe that this should be possible with the increased availability of the astrophysical data. However, we obtain some bounds on the quantum gravity scale following from given realizations. And it was show that our framework makes it possible to retain that the Planck scale is the quantum gravity scale.
The above arguments suggest that the $\kappa$-Minkowski algebra related with the $\kappa$ -Poincar\'{e} Hopf algebra might be able to capture certain aspects of the physics at the Planck scale, which is compatible with the
claim that noncommutative geometry arises from a combined analysis of
special relativity and the quantum uncertainty principle \cite{sergio1,sergio2}.
It is well known that the $\kappa$-Minkowski spacetime leads to a modification
of particle statistics and to deformed oscillator algebras \cite{s4,s5,KS1,KS2,KS3,KS4,KS5}. The twisted
statistics has been used to put bounds on the $\kappa$ parameter within the
context of atomic physics \cite{siva1} and it would be interesting to
compare the bounds arising from other physical scenarios.

\acknowledgments
One of us (A.B.) has been supported by MNiSW Grant No. NN202 318534.
and the Bogliubov-Infeld Program. KSG would like to thank S. Sarkar for discussions.
S.M. acknowledges the support by the Croatian Ministry of Science, Education and Sports grant No. 098-0000000-2865.

\end{document}